\documentclass[11pt,twoside]{article}
\newcommand\gsim{\,\lower3pt\hbox{$\sim$}\llap{\raise2pt\hbox{$>$}}\,}
\newcommand\lsim{\,\lower3pt\hbox{$\sim$}\llap{\raise2pt\hbox{$<$}}\,}

\usepackage{asp2006}
\usepackage{epsf}
\usepackage{psfig}
\usepackage{lscape}

\begin{document}
\title{Modeling the Subsurface Evolution of Active Region Flux Tubes}   
\author{Y. Fan}   
\affil{High Altitude Observatory, National Center for Atmospheric Research, 3080 Center Green Dr., Boulder, CO 80301}    

\begin{abstract} 
I present results from a set of 3D spherical-shell MHD simulations of the buoyant rise of
active region flux tubes in the solar interior which put new constraints on the initial twist
of the subsurface tubes in order for them to emerge with tilt angles consistent with
the observed Joy's law for the mean tilt of solar active regions.
Due to the asymmetric stretching of the $\Omega$-shaped tube by the Coriolis force, a field
strength asymmetry develops with the leading side having a greater field strength and thus being
more cohesive compared to the following side.
Furthermore, the magnetic flux in the leading leg shows more coherent values of local twist
$\alpha \equiv {\bf J} \cdot {\bf B} / B^2$, whereas the values in the following leg show
large fluctuations and are of mixed signs.
\end{abstract}

\section{Introduction}

If we believe that active regions on the solar surface originate from
a strong, predominantly toroidal magnetic magnetic field generated at the base of the
solar convection zone, then we need to explain how the magnetic flux tubes rise through
the solar convection zone to the surface. Despite the turbulent nature of solar
convection, bipolar active regions on the solar surface show remarkable order and
organization as described by the Hale polarity law and Joy's law of active region tilts.
This suggests that the field strength of the rising flux tubes are significantly
super-equipartition compared to the kinetic energy of solar convection in the bulk of the
solar zone such that emerging flux tubes are not severely distorted by the convective
motion and roughly preserve the original orientation of the toroidal magnetic fields
at the base of the solar convection zone.

Previously a large body of calculations based on a highly simplified {\it thin flux tube}
model \citep{spruit1981} has provided important insights into the dynamic evolution of
a $\Omega$-shaped rising flux tube in the solar convective envelope
\citep[e.g][]{dsilva_choud1993,fanetal1993,caligarietal1995,caligarietal1998,
fan_fisher1996}.
These calculations suggest that the field strength of the toroidal magnetic field at the
base of the solar convection zone is in the range of about $3 \times 10^4$ G to $10^5$ G
in order for the emerging tubes to be consistent with the observed properties of solar
active regions.  It is found that the Coriolis force acting on the rising $\Omega$-shaped
tube can produce asymmetries between the leading and following legs of the $\Omega$-tube
which provide explanations for several observed asymmetric properties between the leading
and following polarities of bipolar active regions.  These include (1) a slight tilt of the
emerging loop consistent with the mean tilt of bipolar active regions; 
(2) the asymmetric geometric shape of the emerging loop which may give rise to the
observed asymmetric horizontal motions of the two polarities of an emerging active region, and
(3) the asymmetry in the field strength between the two sides of the loop with a stronger field
along the leading side which may be the cause
of the observed asymmetric morphology for the leading and following polarities.
Since the thin flux tube model satisfies the frozen-in condition perfectly, it faithfully
describes the evolution of the magnetic field strength, the magnetic buoyancy, and hence the
velocity of the rising flux tube without being subject to numerical diffusion.  
\begin{figure}
\centerline{\epsfxsize=4.5 true in \epsfbox{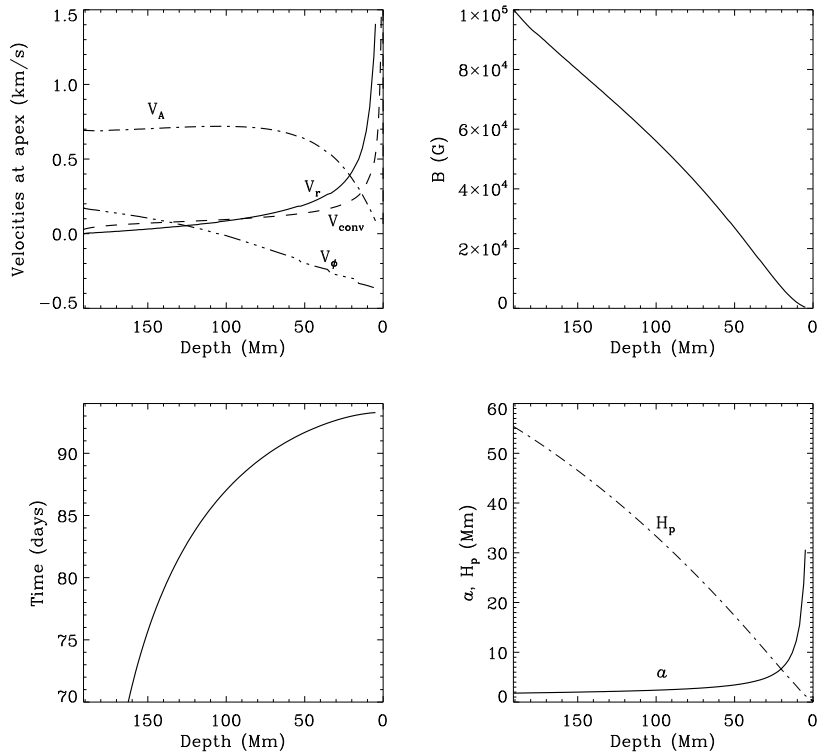}}
\caption{The evolution at the tube apex of ({\it top-left panel}) the Alfv\'en velocity $v_{A}$,
rise velocity $v_r$, convective velocity $v_{\rm conv}$ based on JCD's solar
interior model \citep{jcd1982}, the azimuthal velocity $v_{\phi}$,
({\it top-right panel}) the magnetic field strength $B$, ({\it bottom-left panel}) time
elapsed since the onset of the Parker instability,
and ({\it bottom-right panel}) the tube radius $a$ and the local pressure
scale height $H_p$, as a function of depth, resulting from a thin flux
tube simulation of an emerging $\Omega$-shaped tube described in
\citet[][corresponding to the case shown in the top panel of Figure 1 in that paper]{fan_gong2000}}
\end{figure}
Figure 1 shows the evolution of various quantities at the apex of the rising tube as it
traverse the convection zone, resulting from a thin flux tube simulation of an emerging
$\Omega$-shaped tube developed due to the Parker
instability of a $10^5$ G toroidal flux tube initially in mechanical equilibrium at the base of
the solar convection zone at an initial latitude of $15^{\circ}$ \citep[see][for details
of the simulation]{fan_gong2000}.  It can be seen that the rise velocity
$v_r$ remains quite small ($\lsim 200 {\rm m s}^{-1}$) and
the Alfv\'en speed $v_a$ remains nearly constant (being much greater than both the rise and
the convective flow speed) in the bulk of the
convection zone, until the top few tens of Mm of the convection zone, where $v_r$ accelerates
steeply and $v_a$ decreases rapidly due to the steep super-adiabaticity in this top layer.
At a depth of roughly $20$ Mm, the radius of the tube $a$ exceeds the local pressure scale
height $H_p$ and $v_r$ also exceeds $v_a$.  At this point, the thin flux tube
approximation breaks down and the tube is likely to be severely distorted and fragmented.  
Nevertheless, if one continues to use the $v_r$ from this point on as an estimate, one finds
that the tube will rise through the last 20 Mm depth of the convection zone in a time of
about only 7 hours.
These results provide some basic information for local helioseismology
to estimate possible helioseismic signatures (e.g. acoustic travel time perturbations) for
detecting subsurface emerging active region flux tubes.

Later MHD simulations \citep[e.g][]{emonet_moreno1998, abbettetal2000} of
buoyantly rising flux tubes in local Cartesian geometries
suggested that in order for the flux tube to rise cohesively, a minimum twist of the flux tube
field lines given by $q \equiv B_{\theta} / (r B_z) = (H_p a)^{-1/2}$
is needed \citep{longcopeetal1999},
where $q$ corresponds to the angular rate of field line rotation about the axis over a unit
length along the tube, $B_z$ and $B_{\theta}$ denote respectively the axial and azimuthal
magnetic field in the cylindrical flux tube, $r$ is the distance to the tube axis, $H_p$ is
the local pressure scale height, and $a$ is the tube radius. This minimum twist is found to
be about an order of magnitude too big compared to the twist deduced from vector magnetic
field observations of solar active regions on the photosphere \citep{longcopeetal1999}.
More recently, \citet{fan2008} has carried out a set of 3D anelastic MHD simulations of the
buoyant rise of active region scale flux tubes in the solar interior in a rotating
spherical shell geometry.  These simulations put further constraints on the twist of the
interior flux tubes in order for them to emerge with tilts that are consistent with the observed
mean tilt angles of solar active regions. Consistent with previous thin flux tube calculations
\citep{fanetal1993}, it is found that due to the asymmetric stretching of the flux tube by the
Coriolis force, the $\Omega$-shaped emerging tube develops a field strength asymmetry with
the leading side having a stronger field, and thus being less fragmented compared to the
following side.  We will review these results in the following and further examine the
asymmetry in the field line twist and magnetic helicity transport along the leading and
following side of the emerging tube.

\section{Results from 3D anelastic spherical-shell simulations:}
\citet{fan2008} performed a set of 3D anelastic MHD simulations of the buoyant rise of
magnetic flux tubes in a rotating spherical shell domain given by: 
$r=[r_c, r_t]$, where $r_c = 0.722 R_{\odot}$ is the
base of the convection zone and $r_t = 0.977 R_{\odot}$ is at about
$16 {\rm Mm}$ below the photosphere, $\theta=[0,\pi/2]$,
and $\phi = [0,\pi/2]$. The domain is resolved by a grid with 256 grid points in $r$,
512 grid points in $\theta$, and 512 grid point in $\phi$.  The grid is uniform in
$r$ and $\phi$, but nonuniform in $\theta$, with finer grid spacing in the region
traversed by the rising flux tube and coarser grid in the polar region.
JCD's solar model \citep{jcd1982} is used for the reference
stratification in the simulation domain, and the superadiabaticity in the convection zone
is set to zero, so that the fluid is marginally stable to convection.
Thus we are modeling the dynamic rise of buoyant flux tubes in a quiescent model
convective envelope without the effects of the convective flows. Initially a twisted
toroidal flux tube with an axial field strength of $10^5$ G and a
radius of $a=5.59 \times 10^8 {\rm cm} \approx 0.1 H_p$ is placed
just slightly above the base of the model convective envelope.
For flux tubes with such a strong initial field strength, which is 10 times the field
that is in equipartition with convection, the dynamic effects of convective flows
on the buoyant flux tube is expected to be small \citep[e.g.][]{fanetal2003}
and neglecting convection in the simulations is expected to be a good approximation.
An initial sinusoidal variation of entropy is imposed along the toroidal tube
such that the middle of the toroidal tube is approximately in thermal equilibrium
with the surrounding and thus is buoyant and the two ends (at $\phi=0$ and $\phi=\pi/2$)
of the toroidal tube are in neutral buoyancy.
Thus an $\Omega$-shaped rising flux tube develops due to the initial buoyancy specification.

Simulations where the initial twist of the field lines is set to the critical value needed
for a cohesive rise of the tube, $q = \pm 0.3 a^{-1} \approx \pm (H_p a)^{-1/2}$, show that
the final tilt of the $\Omega$-tube at the apex is largely determined by the sign of the
twist, with positive (negative) $q$ resulting in a clockwise (counter-clockwise) tilt.
Here $q$ denotes the angle of field line rotation about the axis over a unit distance along
the tube.
In these cases, it is found that the writhing of the tube due to the field line twist
dominates the effect of the Coriolis force and produces the wrong sign of tilt at the apex
of the emerging tube, if the observed hemispheric preference of the sign of twist
(i.e. left-handed-twist or negative $q$ in the northern hemisphere) is assumed.
It is found that in order for the emerging tube to show
the correct tilt direction (consistent with observations), the initial twist of
the flux tube needs to be less than a half of that needed for a cohesive
rise. Under such conditions, severe flux loss is found during the rise.
\begin{figure}
\centerline{\epsfxsize=5 true in \epsfbox{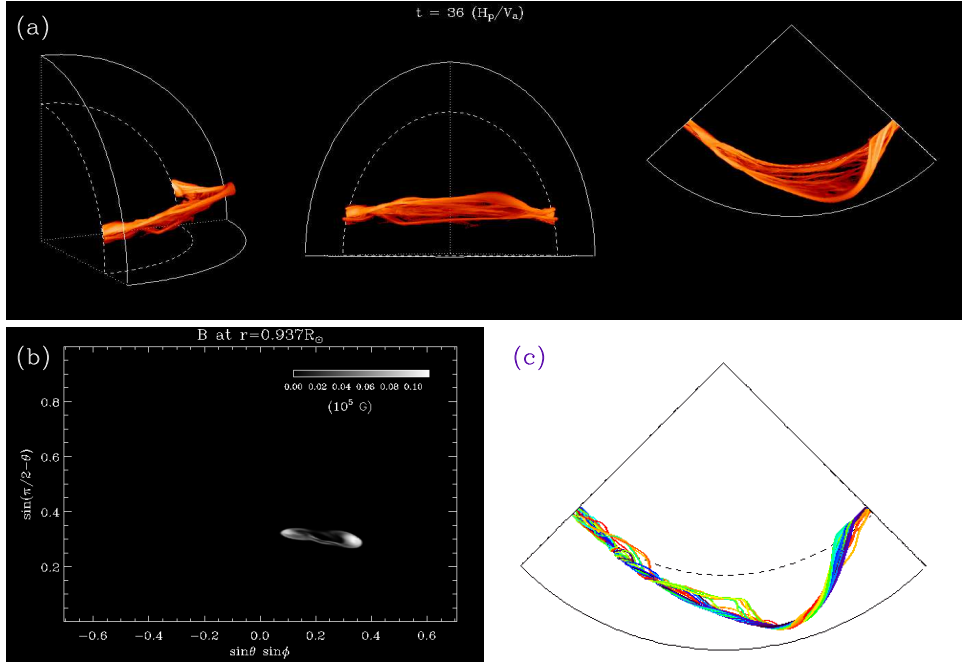}}
\caption{(a) 3D volume rendering of the magnetic field strength of an rising $\Omega$-tube
whose apex is approaching the top boundary, resulting from a simulation with an initial
twist $q = -0.15 a^{-1}$, initial field strength $10^5$ G, and an initial latitude of
$15^{\circ}$ \citep[see the LNT run in][]{fan2008}; (b) A cross section of $B$ near the top
boundary at $r=0.937 R_{\odot}$; (c) selected field lines threading through the coherent apex
cross-section of the $\Omega$-tube.}
\end{figure}
\begin{figure}
\centerline{\epsfxsize=5 true in \epsfbox{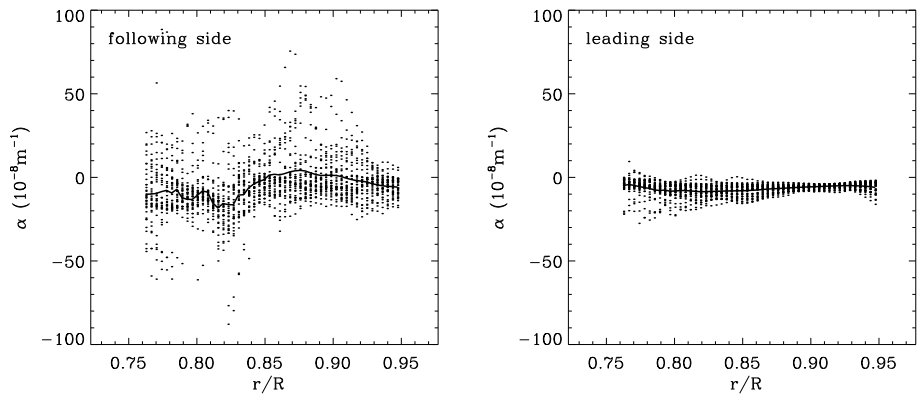}}
\caption{Dots show values of $\alpha \equiv {\bf J} \cdot {\bf B} / B^2$ computed
along each of the selected field lines of the final $Omega$-tube shown in
Figure 2(c) as a function of depth for the following side (left panel) and the leading side
(right panel).  The field line averaged mean $\alpha$ is shown as the solid curve.}
\end{figure}
Figure 2(a) shows the 3D volume rendering of the magnetic field strength of an $\Omega$-shaped
rising flux tube whose apex is approaching the top boundary, resulting from a simulation with
an initial twist $q = -0.15 a^{-1}$, initial field strength of $10^5$ G, starting from an initial
latitude of $15^{\circ}$.  Only about 45\% of the initial flux remains in the coherent portion
of the tube that rises to the upper boundary.  In this case the final $\Omega$-tube
attains a clockwise tilt angle of about $6.5^{\circ}$ for its apex portion (as can be seen in
the cross-section shown in Figure 2(b)), consistent with the mean tilt of solar active regions.
Furthermore, the Coriolis
force drives a retrograde flow along the apex portion, resulting in a relatively greater
stretching of the field lines and hence stronger field strength in the leading leg of the tube.
With a greater field strength, the leading leg is more buoyant with a greater rise
velocity, and remains more cohesive compared to the following leg.
Figure 2(c) shows selected field lines threading through the coherent apex cross-section of the
rising $\Omega$-tube.  One can clearly see the asymmetry where the field lines in the leading
side are winding about each other smoothly in a coherent fashion, while the field lines in
the following side are fraying out.
We have also evaluated the local twist rate given by $\alpha
\equiv {\bf J} \cdot {\bf B} / B^2$, where $J$ is the electric current density, along
each of the selected field lines shown in Figure 2(c) and plotted them (as dots) as a function of
depth in Figure 3 for the following side (left panel) and the leading side (right panel)
respectively.  The mean value averaged over the field lines as a function of depth
is shown as the solid curves.
We can see that the $\alpha$ values along the field lines on the following side
show large fluctuations and have mixed signs, whereas the values along the field lines on
the leading side are more coherent and are consistently negative.
However the mean $\alpha$ averaged over the field lines does not show a significant
systematic difference between the leading and following sides.

\section{Conclusions}
Our 3D simulations of buoyantly rising twisted flux tubes in a rotating spherical
shell geometry suggest a new upper limit for the twist of the subsurface flux tubes
in order for them to emerge with tilt angles consistent with the observed tilts
of the majority of solar active regions. This upper limit is about a half
of the twist needed for a flux tube with the magnetic buoyancy to rise cohesively. 
The simulation results also show that due to the Coriolis force, the leading leg of the
$\Omega$-shaped tube tends to have a stronger field strength, being more buoyant
and more cohesive compared to the following leg.  Furthermore the field lines in the leading
leg show more coherent values of local twist as measured by $\alpha \equiv {\bf J} \cdot {\bf B}
/ B^2$, whereas the field lines in the following leg are more frayed and show larger
fluctuations and mixed signs of twist.
Recent observational studies by \citet{tian_alex2008} have found an systematically greater
helicity injection flux in the leading polarity than the following in several emerging active
regions.
Although in our model, the field lines of the leading leg of the $\Omega$-tube do not show a
systematically greater twist on average, the greater buoyancy and hence higher rise velocity
of the leading leg in conjunction with its more coherent twist may give rise to a greater
upward helicity flux in the leading polarity
comparing to the following as a result of the emergence of the $\Omega$-tube.

\acknowledgements I thank Lirong Tian and David Alexander for helpful discussions.
This work is supported in part by NASA Heliophysics Guest Investigator Grant NNG07EK66I.
NCAR is sponsored by National Science Foundation.

\end{document}